\documentclass[twocolumn]{aastex631}

\begin{document}

\title{Magnetic field monitoring of four massive A-F supergiants\footnote{This work is based on observations obtained at the Canada-France-Hawaii Telescope (CFHT) which is operated by the National Research Council (NRC) of Canada, the Institut National des Sciences de l'Univers of the Centre National de la Recherche Scientifique (CNRS) of France, and the University of Hawaii. The observations at the CFHT were performed with care and respect from the summit of Maunakea which is a significant cultural and historic site.}}

\author{Gregg A. Wade}
\affiliation{Queen's University\\
Department of Physics, Engineering Physics, and Astronomy,
64 Bader Lane
Kingston, ON K7L 3N6
Canada}
\affiliation{Royal Military College of Canada \\
Department of Physics \& Space Science, 
PO Box 17000, Station Forces,
Kingston, ON, Canada K0H1S0}


\author{Mary Oksala}
\affiliation{California Lutheran University\\
Department of Physics, 60 W. Olsen Road, Thousand Oaks, CA, 91360, USA}
\affiliation{LIRA, Paris Observatory\\
PSL University, CNRS, Sorbonne University, Universit\'e Paris Cit\'e, CY Cergy Paris University, 92190 Meudon, France}

\author{Coralie Neiner}
\affiliation{LIRA, Paris Observatory\\
PSL University, CNRS, Sorbonne University, Universit\'e Paris Cit\'e, CY Cergy Paris University, 92190 Meudon, France}

\author{\'Etienne Boucher}
\affiliation{Royal Military College of Canada \\
Department of Physics \& Space Science, 
PO Box 17000, Station Forces,
Kingston, ON, Canada K0H1S0}

\author{James A. Barron}
\affiliation{Queen's University\\
Department of Physics, Engineering Physics, and Astronomy,
64 Bader Lane
Kingston, ON K7L 3N6
Canada}
\affiliation{Royal Military College of Canada \\
Department of Physics \& Space Science, 
PO Box 17000, Station Forces,
Kingston, ON, Canada K0H1S0}



\begin{abstract}

We report magnetic field measurements spanning about 15 years of four massive ($7.5-15~M_\odot$) supergiant stars: $\alpha$~Per (HD\,20902, F5Iab), $\alpha$~Lep (HD\,36673A, F0Ib), $\eta$~Leo (HD\,87737, A0Ib) and 13~Mon (HD\,46300, A1Ib). For each star, spectropolarimetric observations were collected using ESPaDOnS at the Canada-France-Hawaii Telescope. The observed spectra were co-added, normalized, then processed using Least Squares Deconvolution (LSD) to yield mean Stokes $I$ and $V$ profiles. Each spectrum was analyzed to infer the False Alarm Probability of signal detection, and the longitudinal magnetic field was measured. This process yielded persistent detection of magnetic fields in all four stars. The median $1\sigma$ longitudinal field uncertainty of the Zeeman detections was 0.6~G. The maximum unsigned longitudinal magnetic fields measured from the detections are rather weak, ranging from $0.34\pm 0.19$~G for $\alpha$~Lep to $2.61\pm 0.55$~G for 13~Mon. The Zeeman signatures show different levels of complexity; those of the two hotter stars are relatively simple, while those of the two cooler stars are more complex. The stars also exhibited different levels of variability of their Zeeman signatures and longitudinal fields. We report periodic variability of the longitudinal field and (complex) Stokes $V$ profiles of $\alpha$~Per with a period of either 50.75 or 90 days. The (simple) Stokes $V$ profiles of 13~Mon, and probably those of $\eta$~Leo, show global polarity changes once during the period of observation, but the data are insufficient to place strong constraints on the variability timescales. 

\end{abstract}




\section{Introduction}\label{intro}

The era of high-resolution stellar spectropolarimetric magnetometry - which began with the introduction of the MuSiCoS spectropolarimeter \citep{1999A&AS..134..149D} at Pic du Midi Observatory - has led to an explosion of knowledge about stellar magnetism. The past 25 years have witnessed the detection and characterization of magnetic fields in non-degenerate stars spanning the Hertzsprung-Russell (HR) diagram, from the coolest to the hottest stars on the main sequence (MS) and their pre-MS progenitors and post-MS descendants. These investigations underpin a firm conceptual understanding of the systematics of stellar magnetism and connections with underlying stellar structural properties and evolutionary states. 

\begin{table*}
\begin{tabular}{l|c|c|c|c}
\hline	
	&	$\alpha$ Per (HD~20902)	&	$\alpha$~Lep (HD 36673)	&	$\eta$~Leo (HD~87737)	&	13~Mon (HD~46300)	\\
\hline								
Spectral Type	&	F5Iab	&	F0Ib	&	A0Ib	&	A1Ib	\\
T$\textsubscript{eff}$ (K)	&	6350-6580	&	6850-7500	&	$9600\pm 200$	&	$10000\pm 200$	\\
log $\textit{g}$	&	1.76-2.0	&	1.34-2.3	&	$2.05\pm 0.1$	&	$2.15\pm 0.1$	\\
Mass (M$_\odot$)	&	7.43-8.5	&	7.53-13.9	&	8.0 - 11.9	&	8.0 - 15	\\
Radius (R$_\odot$)	&	$60\pm 7$	&	$74\pm 22$	&	47	&	34	\\
v sin i (km s$^{-1}$)	&	19$\pm$1	&	14$\pm$1	&	2	&	0	\\
Source & (1) & (2) & (3) & (3)\\
$P_{\rm rot}^{\rm max}$ (days)	&	160	&	270	&	170	&	130	\\
\hline												
\end{tabular}
    \caption{Characteristics of the four stars studied. Sources of atmospheric/rotational parameters are as follows: (1) \citet{lyubimkov2010} (2) \citet{martin18} (3) \citet{2012A&A...543A..80F}. $P_{\rm rot}^{\rm max}$ is the maximum rotation period as estimated from published $v\sin i$ and radius assuming rigid rotation. }
    \label{tab:Stellar}
\end{table*}

\begin{figure*}
\centering
\includegraphics[width=16cm]{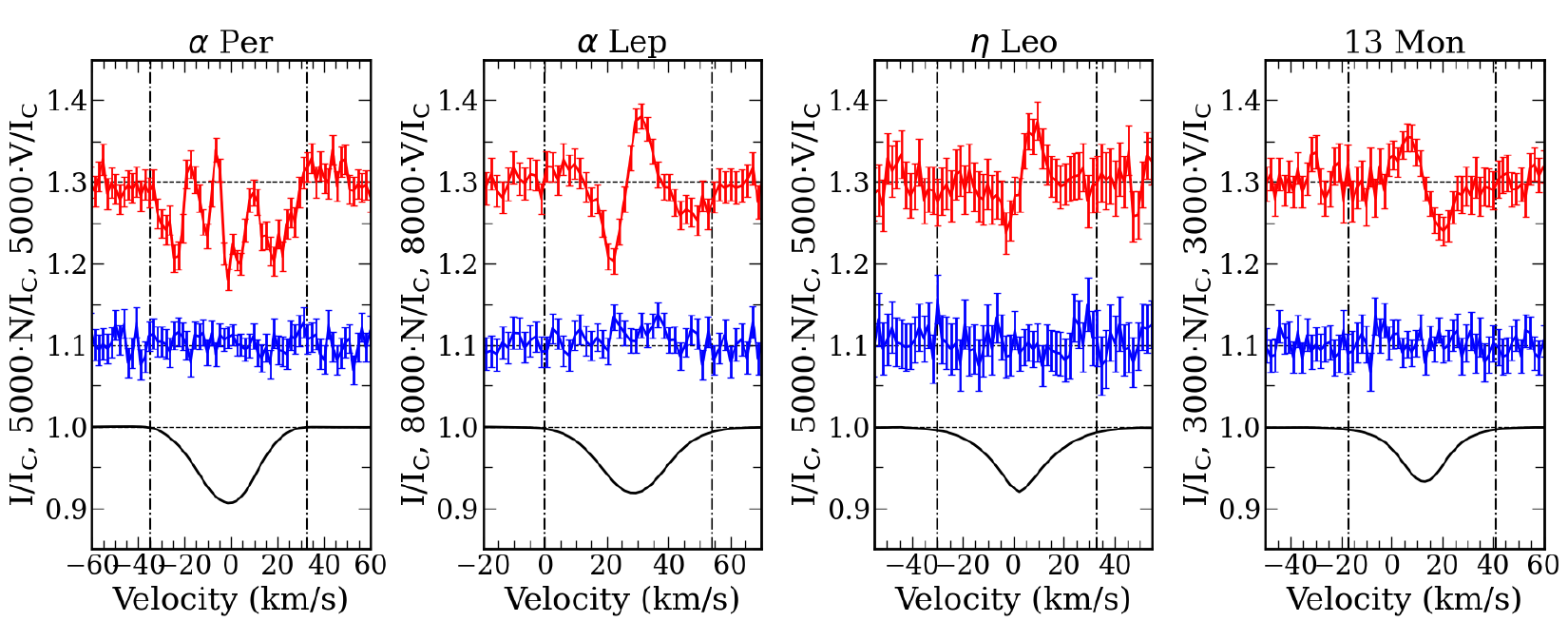}
\centering
\caption{\label{figLSD}Illustrative LSD Stokes $I$ (black, bottom) and $V$ (red, top) profiles, and diagnostic null (blue, middle) profiles, for the 4 stars investigated in this study (with each star indicated above each panel). Dashed lines indicate the  ranges of velocity employed for calculation of the False Alarm Probability and the longitudinal magnetic field. The profiles shown correspond to Night 2 for $\alpha$~Per, Night 3 for $\alpha$~Lep, Night 6 for $\eta$~Leo, and Night 3 for 13~Mon.}
\end{figure*}

In cool stars, characterized by vigorous convective envelopes, ubiquitous magnetic activity 
results from a dynamo powered by the conversion of convective and rotational mechanical energy into magnetic 
energy, generating and sustaining highly structured and variable magnetic fields in their outer envelopes whose 
surface properties correlate strongly with stellar mass, age, and rotation rate \citep[e.g.][]{2011ApJ...743...48W}. These convective dynamos are not fully understood; nevertheless, their basic principles are well established \citep[e.g.][]{2009LRSP....6....4F,2014ARA&A..52..251C}, especially for the main sequence stars and late pre-main sequence stars that have been a focus of observations. 

Field complexity and variability are clear functions of bulk internal stellar structure. For example, main sequence stars like the Sun, with moderately deep surface convection zones, exhibit weak global dipoles, strong small-scale fields, and frequently cyclical variability on timescales of years to decades \citep[e.g.][]{Folsom_2016,Bellotti_2025}. On the other hand, fully convective main sequence M dwarfs and pre-main sequence T Tauri stars largely exhibit strong, stable, large scale fields \citep[e.g.][]{2012EAS....57..165M,2017NatAs...1E.184S}. Many magnetically-active cool stars also exhibit observational evidence of the presence of hot, extended atmospheres \citep[i.e. chromospheres and coronae;][]{1980ARA&A..18..439L}.

Strong magnetic fields are also frequently detected at the surfaces of stars with radiative envelopes, including 
pre-main sequence (Herbig Ae/Be) stars, hot A, B and O-type main sequence stars, early-type giants and (possibly)
supergiants, and white dwarfs and neutron stars \citep[e.g.][]{2009ARA&A..47..333D}. The characteristics of these fields generally differ fundamentally from those of cool stars: they are intense, long-lived, and organized on global scales, and show no clear correlations with stellar rotational properties. It is now generally accepted that they are not currently generated by dynamos, but rather that they are fossil fields, i.e. remnants of field generated, accumulated, or enhanced during earlier phases of stellar evolution \citep[e.g.][]{2009ARA&A..47..333D}. Fossil fields have now been established to exist at the surfaces of non-degenerate stars with masses ranging from about 1.5 to more than 40 times that of the Sun. Notwithstanding the important differences in stellar structure that occur over this enormous range of mass, the characteristics of the magnetic fields remain quantitatively and qualitatively unchanged \citep{2015ASPC..494...30W}. 

In recent years, a third – apparently distinct – character of magnetic field has been discovered in a small but 
growing number of early A-type main sequence stars \citep[e.g.][]{2009A&A...500L..41L,petit11,2016A&A...586A..97B}. Termed “ultra-weak” or “Vega-like” fields, these fields appear to be weak and complex suggesting dynamos \citep[e.g.][]{petit10}. However, given the primarily radiative nature of the atmospheres of their hosts stars they are not likely (purely) dynamo in nature, and have been hypothesized to be the products of interaction between weaker fossil fields and rotational or convective flows \citep[e.g.][]{auriere07,cantiello09}. These fields may provide important guidance for understanding the {\it magnetic desert} phenomenon – a term coined \citep{lignieres14} to describe the dearth of stars hosting fossil fields weaker than a few hundred gauss, and consequently the relative rarity of fossil fields among A, B, and O-type main sequence stars.

While some classes of evolved cool stars have been sufficiently well studied for a reasonable first assay of their magnetic properties \citep[e.g. cool giants;][]{2015A&A...574A..90A},  others remain essentially unexplored. This is especially true of the blue, yellow, and red supergiants. These phases of post-MS evolution are highly sensitive to a number of physical properties of their OB star progenitors, including metallicity, rotation, and mass loss, the latter two of which are known to be strongly influenced by magnetic fields \citep{langer2012}.


\citet{auriere10} reported the first convincing detection of a magnetic field in a highly-evolved supergiant star: the M supergiant Betelgeuse\footnote{In this paper we exclude substantive discussion of the O9Ib star $\zeta$~Ori A, first identified as magnetic by \citet{bouret08}. While spectroscopically classified as a supergiant, this star is located near the end of the main sequence on the HR diagram ($\log g= 3.25$;\citealt{bouret08}), and is better classified as a moderately evolved main sequence object in the evolutionary sense.} Following closely on the heels of that detection, \citet{grunhut10} presented a survey of weak magnetic fields in a sample of over 30 intermediate-mass and high-mass supergiants ranging in spectral type from late M to early A. They detected Zeeman circular polarization signatures in the line profiles of about one-third of their sample, and concluded that the magnetic fields were likely of dynamo origin. The typical $1\sigma$ uncertainty of the longitudinal magnetic field ($B_\ell$) measurements of the detections was about 0.3~G. The three earliest stars of their sample in which magnetic fields were confidently detected were $\alpha$~Lep (F0Iab) and $\alpha$~Per (F5Ib; both subjects of follow-up in the present paper), and $\eta$~Aql (F6Iab; recently confirmed and investigated in greater detail by \citealt{barron22}). 

To our knowledge, $\alpha$~Lep, $\alpha$~Per, and $\eta$~Aql represented the hottest supergiants for which confident magnetic detections had been reported at that time. Although \cite{2003ASPC..305..364V} had previously used the MuSiCoS spectropolarimeter to search for magnetic fields in a sample of A-type and B-type supergiants, no magnetic fields were detected with typical $B_\ell$\ uncertainties of a few tens of G, and best uncertainties of 4~G (for the A supergiant Deneb). Similarly, \cite{shultz14} searched for magnetic fields in six BA supergiant stars ($\beta$ Ori, 4 Lac, $\eta$ Leo, HR~1040, $\alpha$ Cyg, $\nu$ Cep) using ESPaDOnS.  No magnetic fields were detected, with typical uncertainties smaller than 5~G, and best uncertainties of about 1~G. 

\cite{neiner17} were the first to present detections of magnetic fields in A-type supergiants, reporting positive HARPSpol and ESPaDOnS measurements for $\iota$ Car (A7Ib; a definite detection) and $\upsilon$~Car (HR 3890; A7Ib; a more marginal detection. Uncertainties of $B_\ell$\ were typically $\sim 0.2$~G for $\iota$~Car, and 1~G for $\upsilon$~Car. In the context of the LIFE (Large Impact of Magnetic Fields on Stellar Evolution) project, \cite{martin18} reported ESPaDOnS observations of 15 evolved OBA giant and supergiant stars, detecting two: the A5Ib-II supergiant 19 Aur, and HR~3042 (likely a main sequence or early post-main sequence late Bp star). Notably, no magnetic field was detected by \cite{martin18} in the supergiants $\eta$ Leo (A0Ib) or 13 Mon (A1Ib), each with $B_\ell$\ uncertainties of 0.6 G, and both of which are subjects of the present paper. The LIFE project also led to the detections of additional magnetic stars \citep[e.g.][]{2021mobs.confE..47O}. 

\begin{figure*}
\centering
\includegraphics[width=14cm]{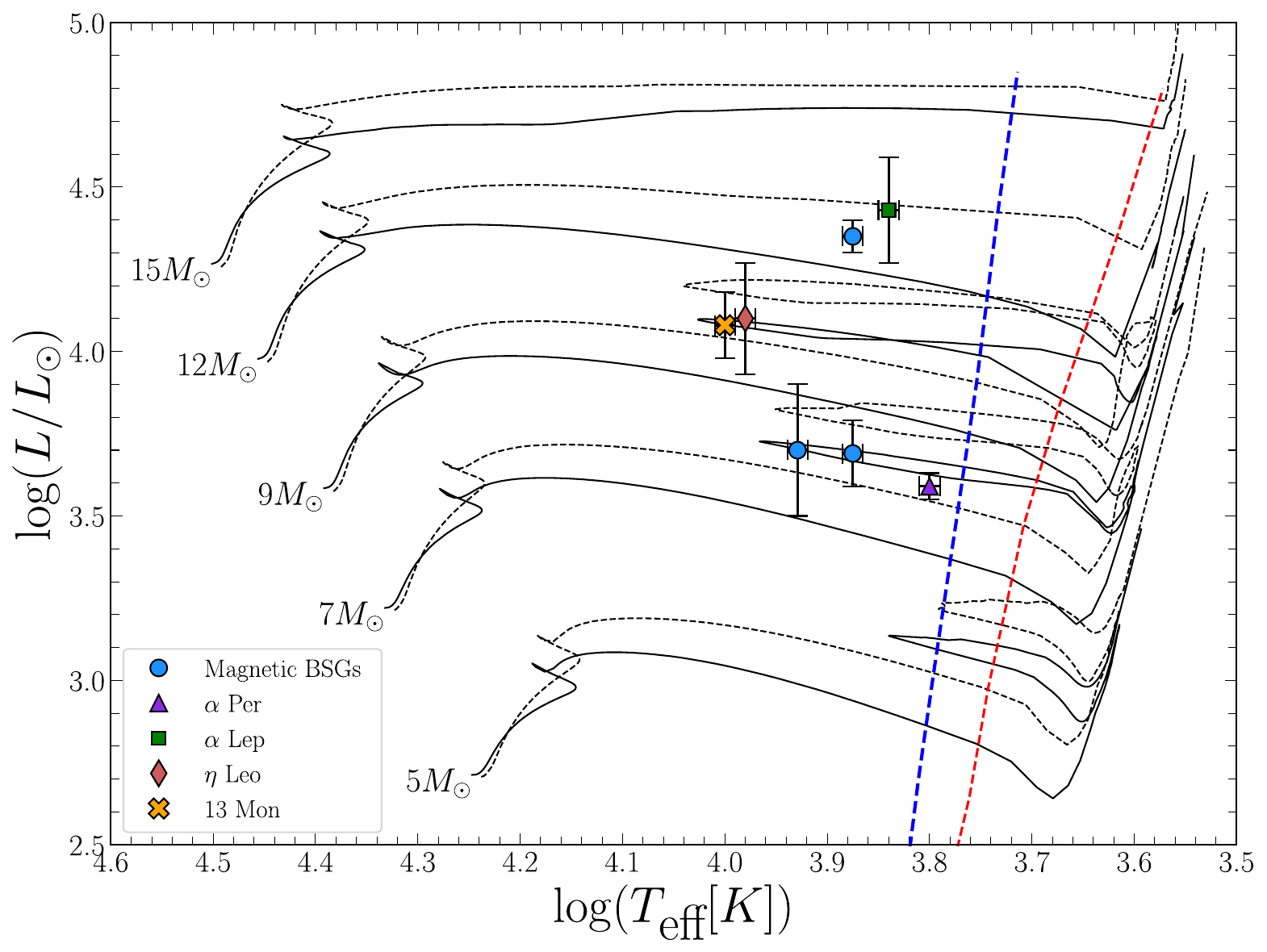}
\centering
\caption{HR diagram showing the inferred positions of the four stars investigated in this study (coloured symbols) along with the positions of other established magnetic early supergiants from the LIFE project (filled blue circles). Evolutionary models are those of \cite{ekstrom12} for $Z=0.014$ with no initial rotation (solid lines) and for initial rotation $\omega=0.5$ (dashed lines), and ignoring the effects of internal and external magnetic fields. Instability strip boundaries are from \cite{Anderson_2016} for first crossing, fundamental Cepheids with initial rotation $\omega=0.5$ and $Z=0.014$.}
\label{Fig:HR diagram}
\end{figure*}

The successful detections of magnetic fields in both cool \citep[e.g.][]{auriere10, grunhut10} and hot \citep{neiner17,martin18} supergiant stars demonstrates that circularly polarized spectra with signal-to-noise ratio (S/N) sufficient to achieve a $B_\ell$\ precision of a few tenths of G are generally necessary in the investigation of their magnetism. In this paper we demonstrate that this is the case, reporting new detections of magnetic fields in two A-type supergiants identified in the context of LIFE, and following up two previously detected F-type supergiants. The paper is structured as follows: In Sects.~\ref{obs} and \ref{sect:mag} we describe the observations and data processing, including preparation of the circularly polarized spectra, extraction of Least-Squares Deconvolution (LSD) profiles, and measurement of the longitudinal magnetic field. In Sect.~\ref{results} we describe the results obtained for each of the 4 stars. In Sect.~\ref{discuss} we discuss the results, situating them in the context of related observational and theoretical results, discuss the likely origin(s) of the fields, and conclude.



\section{Observations and data processing}\label{obs}

In the context of the LIFE program described by \cite{martin18}, we have obtained new spectropolarimetric observations of 4 bright ($1.79<V<4.50$), evolved (luminosity classes Ib/Iab, $\log g\sim 2$), intermediate-to-high-mass ($M\sim 7.5-15~M_\odot$) stars. These targets and their physical properties are summarized in Table~\ref{tab:Stellar}.










The observations were obtained using the ESPaDOnS spectropolarimeter at the Canada-France-Hawaii Telescope (CFHT). ESPaDOnS consists of a high-resolution (resolving power $R\sim 65000$) \'echelle spectrograph which is fibre fed from a Cassegrain-mounted polarimetric module. Each Stokes $V$ observation consisted of a sequence of 4 sub-exposures, between which the polarimetric optics of the instruments were rotated, allowing for the removal of instrumental systematics \citep[see e.g.][]{donati97}. Exposure times were specific to each target, and were typically adjusted to about 90\% of the CCD saturation time according to the official ESPaDOnS exposure time calculator. Sequences of successive Stokes $V$ observations - as many as 40 - were acquired for each star in order to increase sensitivity to weak magnetic fields. The data were reduced using CFHT's Upena pipeline feeding the Libre-ESpRiT reduction code \citep{donati97}. The reduced spectra obtained for each star on each night were co-added, and then normalized to the continuum using polynominal fits to each spectral order. The log of observations is reported in Table~\ref{tab:Ultimate Table}. 

Each spectrum was processed using Least-Squares Deconvolution \citep[LSD;][]{donati97} using the iLSD approach of \citet{kochukhov10}. Line masks were developed using Vienna Atomic Line Database \citep[VALD; e.g.][]{piskunov95} {\tt extract stellar} requests corresponding to the effective temperature and surface gravity of each star (see Table~\ref{tab:Stellar}). Each mask was `cleaned' and `tweaked' \citep{grunhut17} to best match the mean spectrum of each star. This process involved removing lines from the masks produced by, or blended with, broad lines (e.g. H Balmer lines), telluric absorption bands, and instrumental systematics, and adjusting the predicted depths of all other lines to best reproduce their observed depths. The final line masks contained between about 200 lines (for the hottest star, 13 Mon) to over 4000 lines (for the coolest star, $\alpha$~Per), and the extracted LSD profiles consisted of Stokes $I$, $V$, and diagnostic null $N$ pseudo line profiles (Fig.~\ref{figLSD}).


\begin{table*}
    \centering
\begin{tabular}{rcccccccrr}
\hline																	
&Night	&	Date	&	HJD	&	\# &	$t_{\rm exp}$	&	S/N	&	Detection	&	$B_\ell$ 	&	$N_\ell$ 	\\
&	&	M/D/Y	&	(-2450000)	&	Obs	&	(s)	&		&	Prob	&	(G)	&	(G)	\\
\hline
$\alpha$ Per & 1       &       2/6/2009        &       4868.783        &       13      &       10 /40  &       441     &       ND              &   $   1.65   \pm   1.91  $            &       $  1.26  \pm   2.01  $ \\
&2       &       12/4/2009       &       5169.791        &       8       &       15 /60  &       3211    &       \textbf{DD}     &   $   0.09   \pm   0.26  $     &           $  0.26  \pm   0.26  $ \\
&3       &       12/7/2009       &       5172.770        &       8       &       15 /60  &       3566    &       \textbf{DD}     &   $  -0.15   \pm   0.24  $     &           $ -0.33  \pm   0.24  $ \\
&4       &       1/26/2010       &       5222.858        &       8       &       15 /60  &       3414    &       \textbf{DD}     &   $   0.22   \pm   0.25  $     &           $ -0.04  \pm   0.24  $ \\
&5       &       1/28/2010       &       5224.719        &       1       &       15 /60  &       1297&   \textbf{DD}     &           $   1.46   \pm   0.76  $     &           $  0.61  \pm   0.76  $ \\
&6       &       7/24/2010       &       5402.106        &       4       &       15 /60  &       983     &       ND      &           $   0.66   \pm   0.95  $     &           $ -0.05  \pm   0.94  $ \\
&7       &       8/1/2010        &       5410.104        &       4       &       15 /60  &       2494    &       ND      &           $  -0.26   \pm   0.35  $     &           $ -0.28  \pm   0.35  $ \\
&8       &       8/5/2010        &       5414.013        &       4       &       15 /60  &       2053    &       ND      &           $  -0.54   \pm   0.45  $     &           $  0.33  \pm   0.45  $ \\
&9       &       10/16/2010      &       5485.815        &       4       &       15 /60  &       2065    &       \textbf{DD}     &   $   1.50   \pm   0.48  $     &           $ -0.08  \pm   0.47  $ \\
&10      &       10/18/2010      &       5487.843        &       12      &       15 /60  &       3715    &       \textbf{DD}     &   $   1.32   \pm   0.27  $     &           $  0.14  \pm   0.27  $ \\
&11      &       10/19/2010      &       5488.981        &       4       &       15 /60  &       1838    &       \textbf{DD}     &   $   1.27   \pm   0.53  $     &           $ -0.43  \pm   0.53  $ \\
&12      &       11/16/2010      &       5516.841        &       4       &       15 /60  &       2179    &       \textbf{DD}     &   $  -1.22   \pm   0.45  $     &           $  0.60  \pm   0.45  $ \\
&13      &       11/21/2010      &       5521.807        &       4       &       15 /60  &       2116    &       \textbf{DD}     &   $  -0.95   \pm   0.46  $     &           $ -0.36  \pm   0.46  $ \\
&14      &       11/22/2010      &       5522.811        &       4       &       15 /60  &       2777    &       \textbf{DD}     &   $  -0.58   \pm   0.35  $     &           $ -0.15  \pm   0.36  $ \\
&15      &       11/24/2010      &       5524.742        &       4       &       15 /60  &       2603    &       \textbf{DD}     &   $  -0.17   \pm   0.37  $     &           $  0.42  \pm   0.37  $ \\
&16      &       11/25/2010      &       5525.988        &       4       &       15 /60  &       2473    &       \textbf{DD}     &   $  -0.72   \pm   0.39  $     &           $  0.51  \pm   0.39  $ \\
&17      &       11/26/2010      &       5526.937        &       4       &       15 /60  &       2401    &       \textbf{DD}     &   $  -0.33   \pm   0.41  $     &           $  0.44  \pm   0.41  $ \\
&18      &       11/27/2010      &       5527.925        &       4       &       15 /60  &       2525    &       \textbf{DD}     &   $   0.27   \pm   0.39  $     &           $ -0.17  \pm   0.39  $ \\
&19      &       7/4/2011        &       5747.127        &       7       &       3 /12   &       1683    &       \textbf{DD}     &   $   1.28   \pm   0.57  $     &           $  0.08  \pm   0.57  $ \\
&20      &       7/6/2011        &       5749.121        &       7       &       3 /12   &       1718    &       \textbf{DD}     &   $   1.28   \pm   0.57  $     &           $  0.17  \pm   0.57  $ \\
&21      &       7/14/2011       &       5757.135        &       3       &       3 /12   &       1718    &       \textbf{DD}     &   $   0.91   \pm   1.15  $     &           $  0.95  \pm   1.16  $ \\
\hline																	
$\alpha$ Lep& 1	&	9/27/2009	&	5102.115	&	3	&	32 /128	&	2369	&	\textbf{MD}	&	-0.19$\pm$0.42	&	-0.04$\pm$0.43	\\
&2	&	10/30/2017	&	8057.072	&	40	&	18 /72	&	6716	&	\textbf{DD}	&	0.15$\pm$0.19	&	0.01$\pm$0.19	\\
&3	&	3/19/2019	&	8561.750	&	40	&	18 /72	&	6602	&	\textbf{DD}	&	0.34$\pm$0.19	&	0.02$\pm$0.18	\\
\hline																	
$\eta$ Leo&1       &       5/5/2009        &       4956.794        &       6       &       35 /140 &       2100    &       ND      &       $ 0.16  \pm   1.64 $ &$-0.65  \pm   1.65$ \\
&2       &       5/8/2009        &       4959.820        &       6       &       35 /140 &       1535    &       ND      &       $ 0.05  \pm   2.33 $ &$-1.84  \pm   2.36$ \\
&3       &       12/7/2009       &       5173.167        &       3       &       93 /372 &       2524    &       ND      &       $ 0.89  \pm   1.28 $ &$ 1.49  \pm   1.30$ \\
&4       &       2/22/2016       &       7440.838        &       5       &       50 /200 &       1300    &       ND      &       $-1.59  \pm   2.77 $ &$-0.40  \pm   2.77$ \\
&5       &       11/7/2017       &       8065.122        &       20      &       50 /200 &       4370    &       ND      &       $-0.09  \pm   0.92 $ &$-0.54  \pm   0.92$ \\ 
&6       &       11/9/2017       &       8067.109        &       20      &       50 /200 &       4873    &       ND      &       $-1.13  \pm   0.80 $ &$ 0.31  \pm   0.81$ \\ 
&7       &       1/1/2018        &       8120.108        &       20      &       50 /200 &       4632    &       \textbf{DD}&    $-1.28  \pm   0.84 $ &$ 0.65  \pm   0.84$ \\ 
&8       &       11/19/2018      &       8442.150        &       9       &       50 /200 &       2698    &       ND      &       $ 1.35  \pm   1.30 $ &$-0.99  \pm   1.30$ \\
&9       &       3/15/2019       &       8557.988        &       20      &       50 /200 &       4389    &       \textbf{MD}&    $-1.94  \pm   0.91 $ &$ 0.30  \pm   0.91$ \\
&10      &       3/22/2019       &       8565.003        &       20      &       50 /200 &       5170    &       \textbf{MD}&    $-1.87  \pm   0.83 $ &$-0.26  \pm   0.80$ \\
&11      &       5/31/2019       &       8634.825        &       20      &       50 /200 &       4533    &       ND      &       $-2.45  \pm   0.77 $ &$-1.09  \pm   0.77$ \\
&12      &       6/8/2019        &       8642.772        &       20      &       50 /200 &       5272    &       ND      &       $-0.33  \pm   0.71 $ &$-0.54  \pm   0.70$ \\
&13      &       6/10/2019       &       8644.773        &       20      &       50 /200 &       5036    &       ND      &       $-1.14  \pm   0.79 $ &$ 0.02  \pm   0.78$ \\
&14 & 1/4/2024 & 10314.158 & 19 & 50 /200 & 3810 & ND & $-0.29\pm 0.97$ & $0.43\pm 0.97$ \\
&15 & 1/19/2024 &  10329.046 & 20 & 50 /200 & 5050 & ND & $0.78\pm 0.76$ & $0.49\pm 0.76$ \\
\hline																	
13 Mon & 1       &       2/19/2016       &       7437.811        &       2       &       138 /552        &       1192    &       ND      &     $-0.81 \pm    1.69$      &   $-0.87  \pm   1.69$        \\
&2       &       11/6/2017       &       8064.021        &       11      &       138 /552        &       3991    &       \textbf{DD}&  $ 2.61 \pm    0.55$      &   $-0.06  \pm   0.55$        \\
&3       &       11/9/2017       &       8067.032        &       11      &       138 /552        &       4299    &       \textbf{DD}&  $ 2.31 \pm    0.62$      &   $ 0.24  \pm   0.62$        \\
&4       &       1/1/2018        &       8119.811        &       11      &       137 /548        &       4087    &       \textbf{DD}&  $-1.23 \pm    0.53$     &    $-0.48  \pm   0.53$       \\
&5       &       1/24/2019       &       8507.806        &       11      &       138 /552        &       3411    &       ND      &     $-1.50 \pm    1.10$      &   $-0.31  \pm   1.03$        \\
&6       &       1/27/2019       &       8510.742        &       11      &       138 /552        &       3721    &       \textbf{MD}&  $-1.30 \pm    0.61$      &   $-0.48  \pm   0.60$        \\ 
&7       &       3/18/2019       &       8560.808        &       11      &       138 /552        &       3868    &       ND      &     $-1.90 \pm    0.80$      &   $ 0.36  \pm   0.79$        \\ 
&8       &       12/06/2020      &       9190.116        &       11      &       138 /552        &       3740    &       ND      &     $-0.80 \pm    0.70$      &   $ 0.22  \pm   0.71$     \\
\hline
\end{tabular}
    \caption{Log of spectropolarimetric observations, including date, HJD, number of co-added observations per night, exposure time, signal-to-noise ratio in the null spectrum measured between 500-505~nm, detection probability, and longitudinal magnetic field measured from both $V$ and $N$ LSD profiles.}
    \label{tab:Ultimate Table}
\end{table*}

\section{Magnetic diagnosis}\label{sect:mag}

\subsection{Detection confidence}

To assess the presence of significant circular polarization indicative of a magnetic field, we calculated the detection probability \citep[e.g.][]{donati97} from each Stokes $V$ LSD profile. The integration limits were determined visually by identifying the limits of the Stokes $I$ line profile of each star \citep[e.g.][]{wade2000}: [-33,26] km/s for $\alpha$~Per, [-1,52] km/s for $\alpha$~Lep, [-34,38] km/s for $\eta$~Leo, and [-9,33] km/s for 13 Mon. As is common convention, a probability greater than 99.999\% is considered a formal Definite Detection (DD), a probability between 99.9\% and 99.999\% is considered a Marginal Detection (MD), and anything less is considered a Non-Detection (ND).\footnote{That said, these thresholds are rather conservative, and Zeeman signatures can often be identified by-eye in profiles corresponding to formal MDs and even NDs.} A similar test of confidence was performed on the LSD $N$ profiles. While our LSD Stokes $V$ profiles resulted in NDs, MDs, and DDs, all of the LSD $N$ profiles yielded NDs. The results of the detection probability analysis are summarized in column 7 of Table~\ref{tab:Ultimate Table}. Illustrative LSD profiles of each star are presented in Fig.~\ref{figLSD}, and the complete datasets corresponding to identifiable detections are illustrated for each target in Figs.~\ref{fig:lpv_aper}, \ref{fig:lpv_alep}, and \ref{fig:lpv_etaleo13mon}.

\subsection{Longitudinal magnetic field}

The mean longitudinal magnetic field ($B_\ell$) is the line-of-sight component of the stellar magnetic field, weighted and integrated over the visible hemisphere of the star. It can be computed via the first-order moment of the Stokes $V$ profile within the line \citep[e.g.][]{donati97,wade00}:

\begin{equation}
B_\ell = -2.14\times10^{11} \frac{\int vV(v)\,dv}{\lambda zc\int [I_c - I(v)]\,dv}
\end{equation}

\noindent where $B_\ell$ is the longitudinal field in gauss, $V(v)$ is the Stokes $V$ profile at velocity $v$ relative to the centre-of-gravity of the line, $I(v)$ is the Stokes $I$ profile, $\lambda$ is the wavelength of the line in nm, $I_{\rm c}$ is Stokes $I$ continuum level, $c$ is the speed of light in km/s, and $z$ is the Land\'e factor used in calculating the LSD profiles. The integration limits described in the previous subsection were used. The longitudinal field was measured from each Stokes $V$ and null $N$ profile. These measurements are reported in the final two columns of Table \ref{tab:Ultimate Table}.

\section{Results}\label{results}





\subsection{\texorpdfstring{$\alpha$}{alpha}~Per}\label{Sect:alphaper}
All modern determinations of the atmospheric parameters of $\alpha$ Per (F5Iab) imply that it is the coolest of the stars in this survey, with $T_{\rm eff}$ between 6300-6600~K and $\log g$ from 1.75-2.0. (See \citealt{lee12} for a good summary.) No GAIA parallax is available for this star, but the Hipparcos parallax ($\pi=6.44\pm 0.17$~mas) is sufficiently precise to allow us to usefully infer the star's luminosity and HR diagram position\footnote{In this work we employ the model evolutionary calculations of \citep{ekstrom12}, specifically those corresponding to no initial rotation and those corresponding to an initial rotation of 50\% critical. These models do not consider the influence of surface or interior magnetic fields.}. 

We adopt the atmospheric parameters determined by \citet{lyubimkov2010}, $T_{\rm eff}=6350\pm 100$~K and $\log g=1.90\pm 0.04$, the Hipparcos parallax, and a bolometric correction (0.01~mag) computed using the relations of \citet{1996ApJ...469..355F} as reported by \citet{2010AJ....140.1158T}. \citet{2014AJ....147..137L} reports an extinction $A_{\rm v}=0.09$~mag. Together these values imply a luminosity $\log L/L_\odot=3.59\pm 0.04$. The HR diagram position, as shown in Fig.~\ref{Fig:HR diagram}, places the star mid-way through the blue loop of a $\sim 6-7~M_\odot$ evolutionary model (depending on assumed initial rotation) close to the blue edge of the Instability Strip. The temperature and luminosity imply a radius of about 50~$R_\odot$, consistent with that observed interferometrically by \citet{2021AJ....162..198B}. Combined with the evolutionary mass, this radius indicates a surface gravity $\log g=1.81$, in good agreement with the measured value.

\begin{figure}
\centering
\includegraphics[width=8cm]{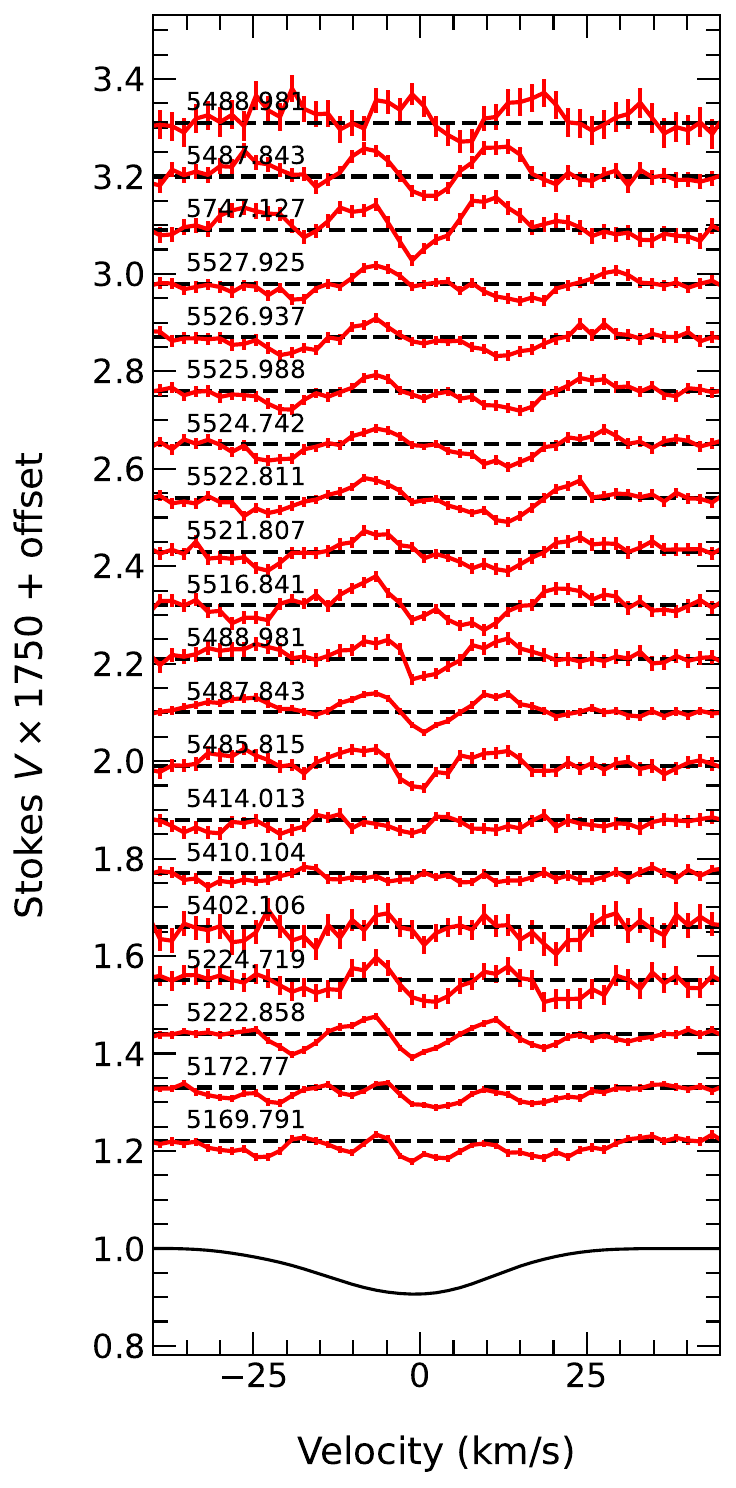}
\centering
\caption{LSD Stokes $V$ profiles of $\alpha$~Per showing significant signal. Profiles are offset vertically in chronological order from bottom to top, with the associated HJD of each profile indicated at left. The Stokes $I$ LSD profile from the first observation is shown at the bottom for context.}
\label{fig:lpv_aper}
\end{figure}




As reported in Table \ref{tab:Ultimate Table}, $\alpha$~Per was the most intensely monitored of our targets, with 21 nights of observations spanning more than two years. After LSD analysis, twenty profiles show Zeeman signatures in Stokes $V$. As illustrated in Fig.~\ref{fig:lpv_aper}, the profiles exhibit very complex, irregular, and variable shapes that change polarity several times across the profile. After the test of detection confidence, 18 of these signatures corresponded to formal DDs, although Zeeman signatures were observed in all profiles.  Integrating over a range of [-33,26]~km/s, the longitudinal field $B_\ell$ varied from $-0.95\pm 0.46$~G to $1.46\pm 0.76$~G. During the most intensely monitored period in 2010, $B_\ell$ was observed to vary systematically and to change sign on timescales of tens of days.

The spectrum of $\alpha$~Per is very slightly variable. No emission is evident in H$\alpha$ or in the cores of the Ca~{\sc ii} H and K lines, classical diagnostics of the presence of chromospheres \citep[e.g.][]{1980ARA&A..18..439L}. These results are consistent with the report of \cite{lee12}. Small variations in the radial velocities and depths of our LSD profiles are apparent. By comparing the profiles of spectra obtained on subsequent (or nearly-subsequent) nights, it is clear that the timescale of these variations is longer than a few days.


\begin{table}
\centering
\begin{tabular}{c|c|c}
\hline
Period&Frequency&Reference\\
(d)&(d$^{-1}$)&\\
\hline
128.2&0.0078&\cite{lee12}\\
87.7&0.0114&\cite{Hatzes_1995}\\
77.717&0.012867&\cite{butler_1998}\\
9.8&0.102&\cite{Hatzes_1995}\\
\hline												
\end{tabular}
    \caption{Literature RV periods and corresponding frequencies of $\alpha$~Per.}
    \label{tab:alphaPer_lit_P}
\end{table}

\begin{figure}
\centering
\includegraphics[width=8cm]{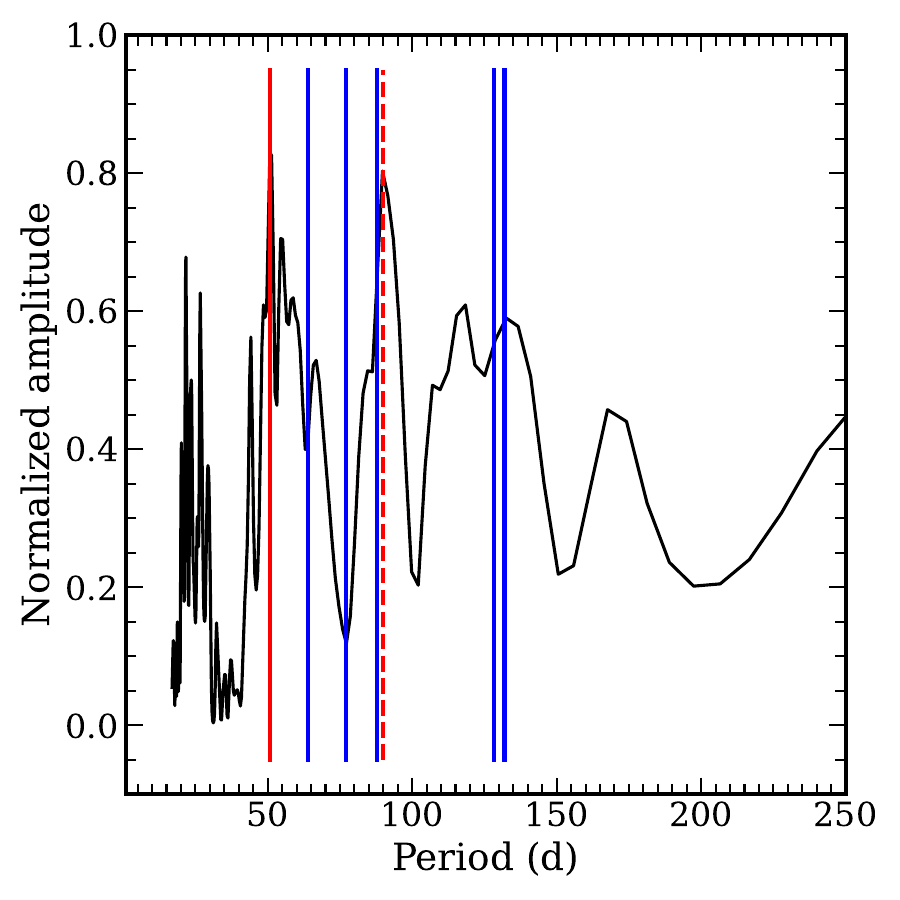}
\centering
\caption{Lomb-Scargle periodogram of the longitudinal magnetic field measurements of $\alpha$~Per presented in Table~\ref{tab:Ultimate Table}. The two most significant periods (50.75~d and 90~d, respectively) are indicated with red solid and dashed lines. Solid blue lines indicate periods reported in the literature and described in Sect.~\ref{Sect:alphaper}.}
\label{fig:periodogram}
\end{figure}

\begin{figure*}
\centering
\includegraphics[width=18cm]{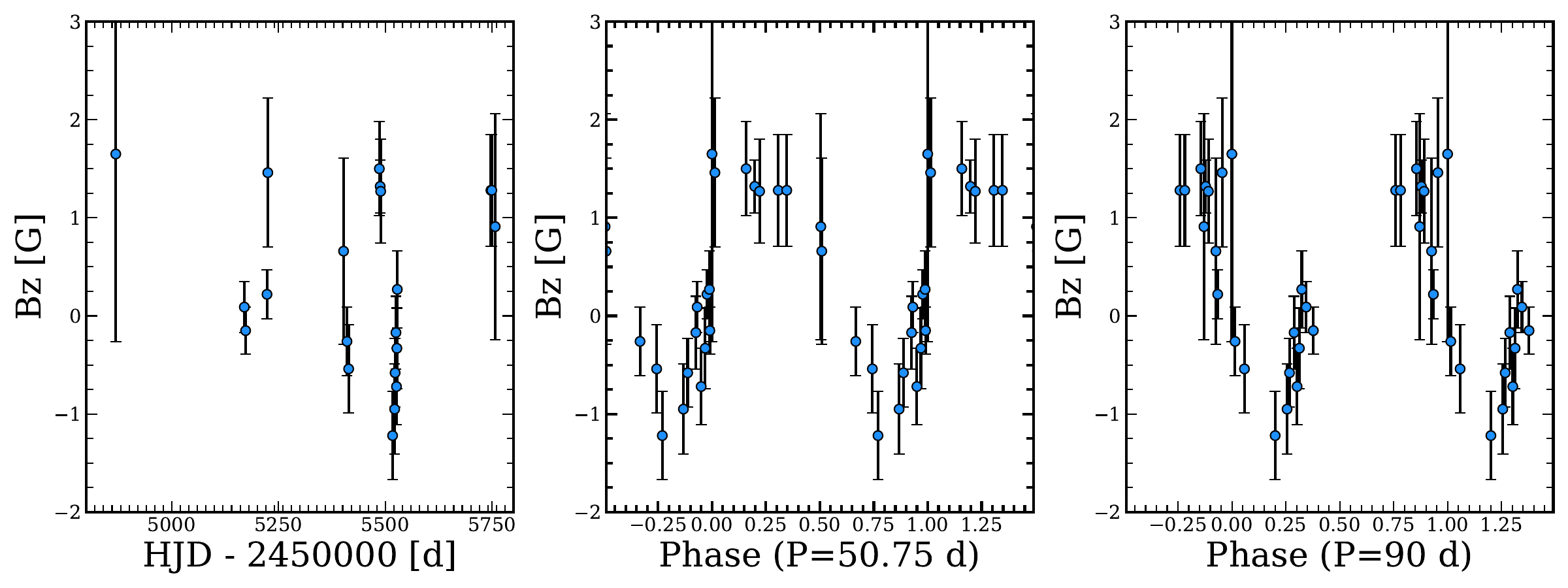}
\centering
\caption{\label{Figbz}Longitudinal magnetic field of $\alpha$~Per (Table~\ref{tab:Ultimate Table}) versus Heliocentric Julian Date (left panel), and folded with the two most significant periods (50.75~d in the middle panel, 90~d on the right) detected in the periodogram (Fig.~\ref{fig:periodogram}.}
\label{fig:LongField}
\end{figure*}

\cite{lee12} investigated the long-term RV variations of $\alpha$~Per, identifying a clear signal at 128.2~d that they speculated was a consequence of rotational modulation. It might reasonably be expected that the longitudinal field of $\alpha$~Per is also rotationally modulated, and would therefore also vary on that timescale. We performed a period search of our timeseries of $B_\ell$ measurements to identify likely periodic signals. We detect significant signals in the periodogram (Fig.~\ref{fig:periodogram}) at 50.75~d and 90~d. Neither of these periods coincides with that of \citet{lee12}. Other periods reported in the literature are summarized in Table~\ref{tab:alphaPer_lit_P}, with the 87.7~d period reported by \citet{Hatzes_1995} being the only one in close agreement with our own. The longitudinal field measurements are shown versus HJD and phased with both of these significant periods in Fig.~\ref{fig:LongField}. 

Either of these periods could plausibly be associated with the rotation of $\alpha$~Per. Adopting the projected rotational velocity of $19$~km/s \citep{lyubimkov2010} and the interferometric radius of $53~R_\odot$ \citep{2021AJ....162..198B}, these periods would imply rotation axis inclinations of $21\degr$ and $40\degr$, respectively. 

The persistent complexity and variability of the Stokes $V$ profiles is reflective of the complex magnetic fields often observed in cool stars with dynamo-driven magnetic fields \citep[e.g.][]{2022A&A...659A..71W,2023MNRAS.525..455D}. The profiles exhibit a qualitative similarity to those of Polaris \citep{barron22}, a somewhat cooler (F8Ib) supergiant (and Cepheid variable) that also exhibits an apparently stable, rotationally modulated longitudinal magnetic field variation (Barron et al., in preparation). Considering profile morphology, variability, and detection incidence, \citet{grunhut10} and \citet{barron22} concluded that the magnetic fields of cool supergiants were of predominantly dynamo origin. $\alpha$~Per appears to fit within this general picture.

Given the strong variability observed on timescales of tens of days, future monitoring should focus on multi-semester, high cadence ($\sim$weekly-monthly) monitoring of this star, which is an excellent candidate for detailed follow-up using Zeeman-Doppler Imaging.

\subsection{\texorpdfstring{$\alpha$}{alpha}~Lep}
 As described by \citet{lyubimkov2010}, the values of the effective temperature and surface gravity of $\alpha$ Lep (F0Ib) reported in the literature span a large range. We adopt the atmospheric parameters determined by \citet{lyubimkov2010} ($T_{\rm eff}=6850$~K, $\log g=1.34$), along with the Gaia DR3 parallax ($\pi=1.68\pm 0.28$~mas) and a bolometric correction (0.02~mag) computed using the relations of \citet{1996ApJ...469..355F} as reported by \citet{2010AJ....140.1158T}. \citet{2014AJ....147..137L} reports an extinction $A_{\rm v}=0.06$~mag, although assuming a somewhat larger distance than we infer. Together these values imply a luminosity $\log L/L_\odot=4.43\pm 0.16$. The HR diagram position (Fig.~\ref{Fig:HR diagram}) places the star on the evolutionary track of a $\sim 12-13~M_\odot$ star. The effective temperature and luminosity imply a radius of about 115~$R_\odot$, indicating a surface gravity of about $\log g=1.42$, again in good agreement with the measured value.

\begin{figure}
\centering
\includegraphics[width=8cm]{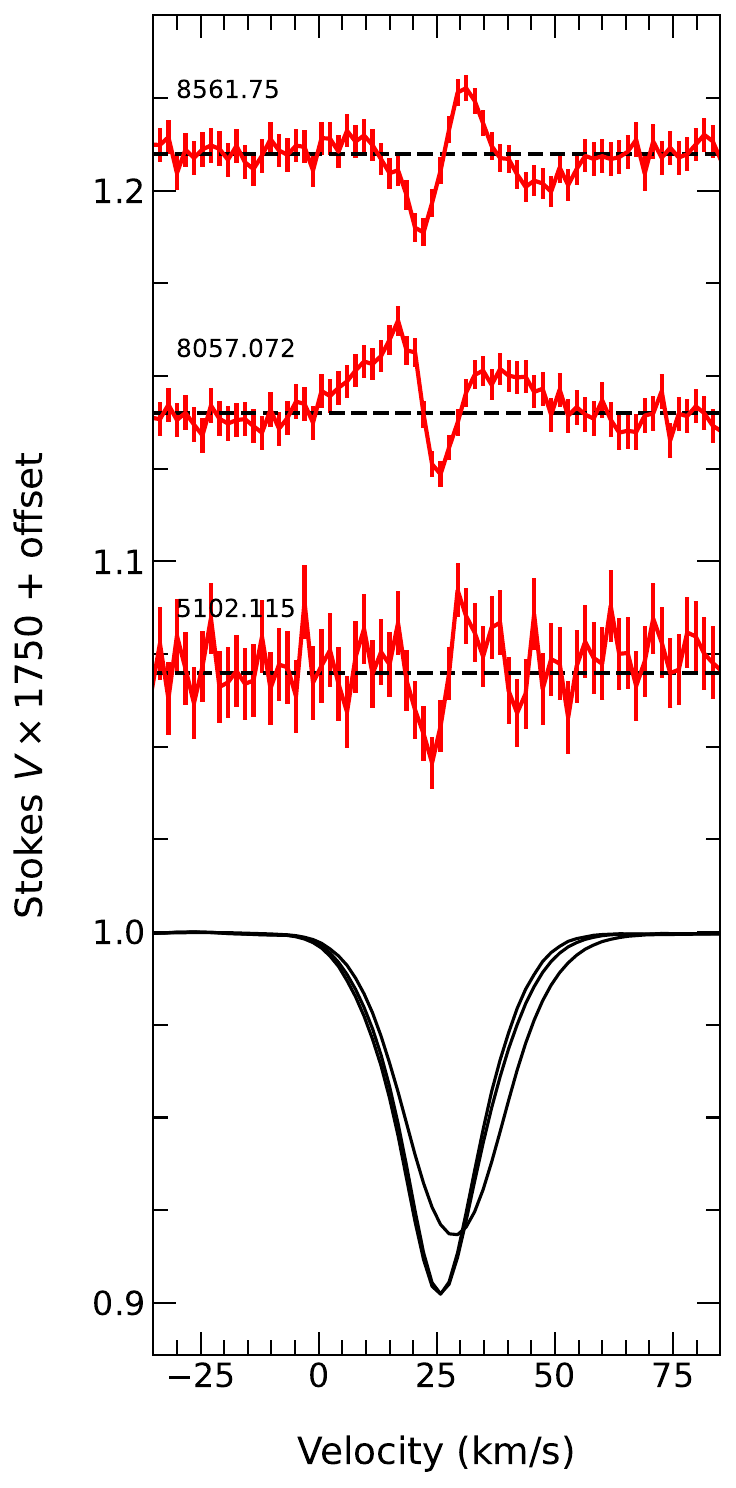}
\centering
\caption{LSD Stokes $V$ profiles of $\alpha$~Lep. Profiles are offset vertically in chronological order from bottom to top, with the associated HJD of each profile indicated at left. The Stokes $I$ LSD profiles of all 3 observations are shown at the bottom. Note the weaker, redshifted profile observed in 2019.}
\label{fig:lpv_alep}
\end{figure}

As summarized in Table \ref{tab:Ultimate Table}, this star had the least number of nights observed in our study, with one night per year in 2009, 2017, and 2019, for a total of three nights. After LSD, all profiles exhibited clear Zeeman signatures in Stokes $V$, corresponding to one MD and two DDs. The latter two profiles exhibit relatively complex shapes that change polarity more than once across the profile (see Figure \ref{fig:lpv_alep}).  While the line profiles observed in 2009 and 2017 are nearly identical, those from 2019 exhibit clearly different shapes. Integrating over range of [-1,52]~km/s, the longitudinal field varied from $-0.19\pm 0.42$~G to $+0.34\pm 0.19$~G. The relatively complex Stokes $V$ morphology and weak longitudinal field again suggest a predominantly dynamo-generated magnetic field, although additional observations are required to form a robust conclusion.

The spectrum of $\alpha$~Lep is somewhat variable, showing especially significant changes in line shape between 2017 and 2019. No emission is evident in H$\alpha$, nor is there obvious evidence of a reversal in the cores of the Ca~{\sc ii} H and K lines. The star is not reported to be a spectroscopic binary \citep{2012AN....333..663S}.

Given the small number of existing observations, future monitoring on a variety of timescales (from $\sim$days to the rotational timescale, likely of order years) should be undertaken.



\subsection{\texorpdfstring{$\eta$}{eta}~Leo}
Historical determinations of the atmospheric parameters of the A0Ib supergiant $\eta$~Leo agree on an effective temperature of $9.5-10$~kK and $\log g$ of about 2.0. The most reliable modern study by \citet{2012A&A...543A..80F} reports $T_{\rm eff}=9.6\pm 0.2$~kK, $\log g=2.05\pm 0.1$, and $v\sin i=2$~km/s, the latter with an uncertainty of 3-5~km/s. 

We adopted the bolometric correction, colour excess, and total-to-selective extinction of \citet{2012A&A...543A..80F}. Using the GAIA DR3 parallax we compute a luminosity $\log L/L_\odot=4.10\pm 0.17$. This value is significantly larger than those computed by \citet{martin18} ($3.77\pm 0.06$) and \citet{2021A&A...650A.128G} ($3.53\pm 0.2$), both of whom employed the (significantly different and less precise) GAIA DR2 parallax. Placing the star on the HR diagram, our parameters imply a mass of about 9~$M_\odot$ according to the evolutionary tracks of \citet{ekstrom12}. As illustrated in Fig.~\ref{Fig:HR diagram}, the star appears to be located near the blueward extremum of the blue loop following the red giant branch (RGB). This HR diagram location implies a radius that is somewhat larger ($\sim 40~R_\odot$) than that inferred from the star's angular diameter \citep[about 28~$R_\odot$;][]{2005A&A...431..773R}, although the predicted surface gravity is reasonably close to the measured value (2.18 vs. 2.05). 

\begin{figure*}
\centering
\includegraphics[width=8cm]{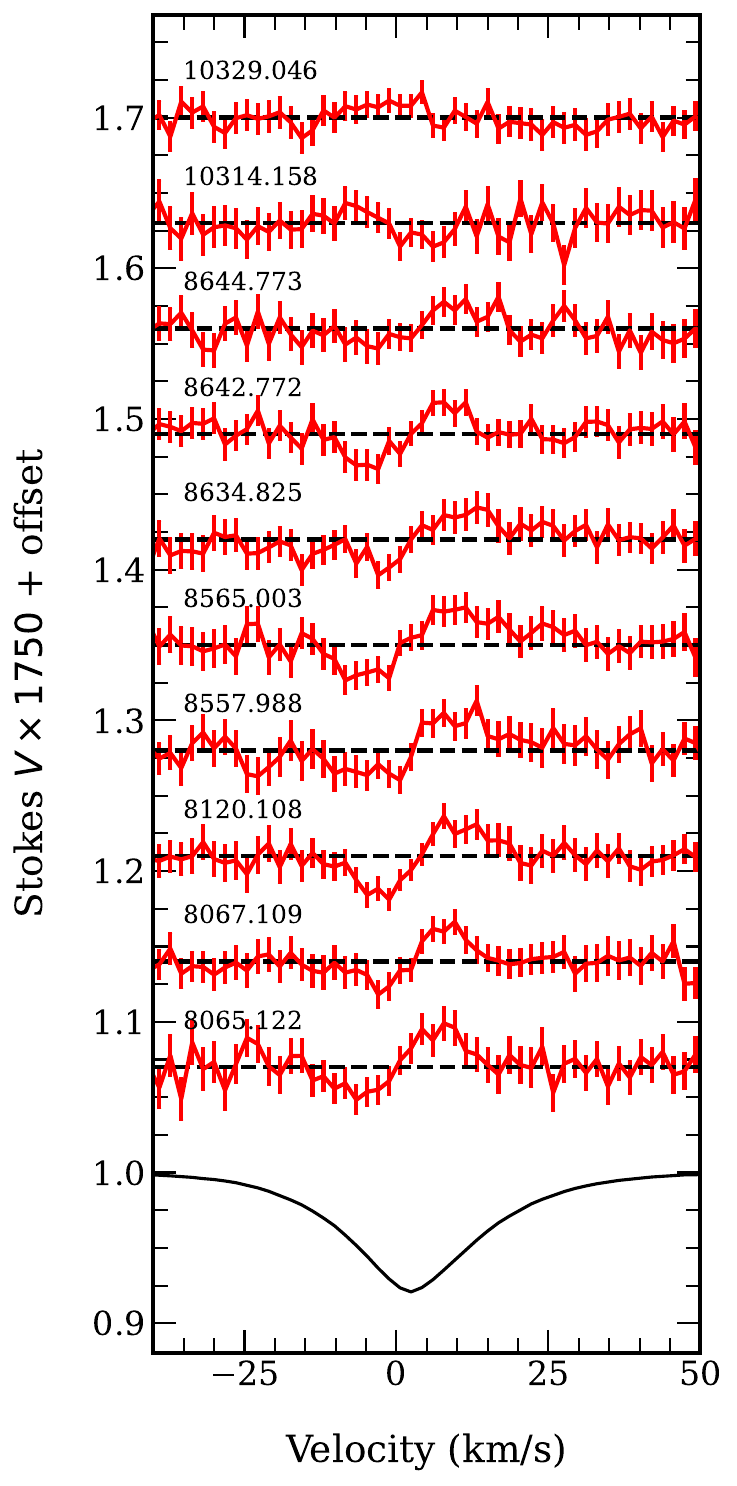}\includegraphics[width=8cm]{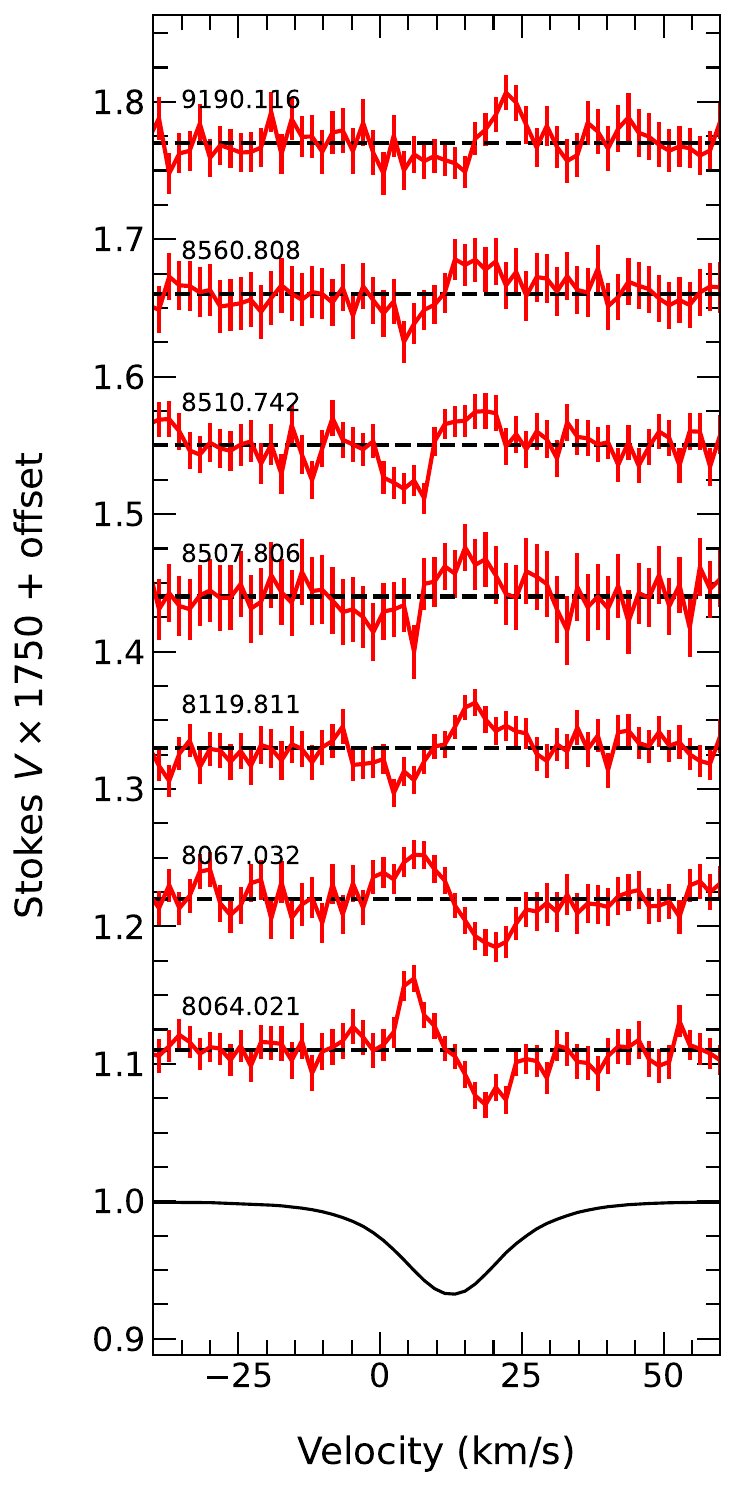}
\centering
\caption{LSD Stokes $V$ profiles of $\eta$~Leo (left) and 13~Mon (right) showing significant signal. Profiles are offset vertically in chronological order from bottom to top, with the associated HJD of each profile indicated at left. The Stokes $I$ LSD profiles corresponding to the first observations are shown at the bottom for context.}
\label{fig:lpv_etaleo13mon}
\end{figure*}

\begin{figure}
\centering
\includegraphics[width=8cm]{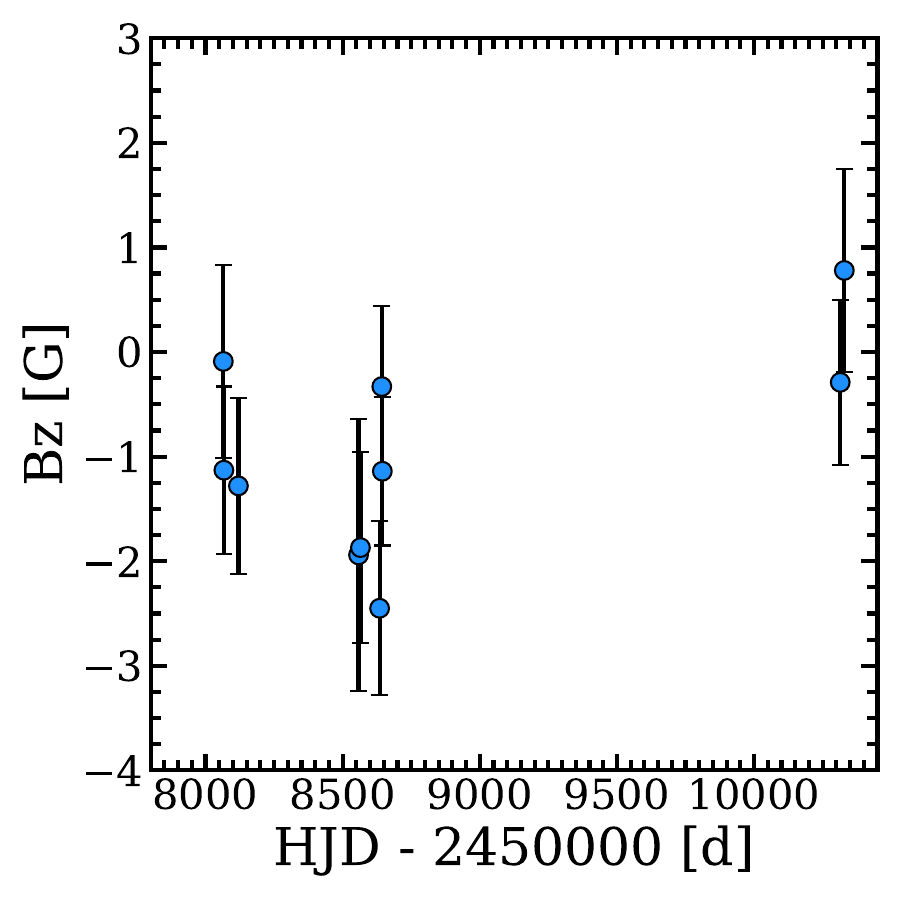}
\centering
\caption{Longitudinal field measurements of $\eta$~Leo versus HJD.}
\label{fig:etaleobz}
\end{figure}

In this study we have collected 15 nights of observations spanning over 14 years. The first observations (all NDs with no evidence of any magnetic signatures) were reported by \citet{shultz14} and \citet{martin18}. We obtained the first obvious evidence of a magnetic field in November 2017. While this measurement still corresponded formally to a ND, a clear Stokes $V$ variation was apparent across the profile, and subsequent observations exhibited the same Stokes $V$ morphology and ultimately led to a DD in January 2018. Following LSD, 10 observations are found to exhibit visually apparent Zeeman signatures, with no significant signal in $N$. The profiles obtained between 2017 and 2019 (see Fig.~\ref{fig:lpv_etaleo13mon}) exhibit a very constant shape, all with the same (negative) polarity. Observations obtained in 2024 appear to have a different shape and amplitude, and may suggest a changing of polarity (based both on profile morphology and a change in sign of the longitudinal field). Integrating over a range of [$-34,+38$] km/s, the longitudinal field measured from the 10 profiles with apparent signatures varies from $-2.45\pm 0.77$~G to $+0.78\pm 0.76$~G (see Fig.~\ref{fig:etaleobz}). At the typical precision of these observations (about 0.8~G) the variation of the longitudinal field is marginally significant. 

The spectrum of $\eta$~Leo is weakly variable over the period of observation. However, no emission is present in H$\alpha$, nor is any present in the cores of the Ca~{\sc ii} H and K lines.

The simple morphology and weak variability of the Zeeman signatures of $\eta$~Leo constrast strongly with those of $\alpha$~Per and $\alpha$~Lep. These properties show some similarities to those of main sequence stars of similarly high effective temperatures, the Ap/Bp stars \citep[e.g.][]{1980ApJS...42..421B}. As described in the introduction, those stars are generally believed to host fossil magnetic fields. If $\eta$~Leo does host a fossil field, and that field represented the remnant of a fossil field that was present on the main sequence, the measured maximum longitudinal field of $|B_\ell|=2.5$~G could imply a flux-conservation-extrapolated ZAMS dipole field strength of at least 1.4~kG.

Continued monitoring at roughly twice-higher precision with a semi-annual cadence for the next $\sim 5$ years is required to confirm the variability, establish its origins, and potentially determine the star's rotation period. 




\subsection{13 Mon}
13 Mon is the hottest star in this survey, with $T_{\rm eff}=10000\pm 200$~K and $\log g=2.15\pm 0.1$ as reported by \citet{2012A&A...543A..80F}. As with $\eta$~Leo, we adopt the atmospheric parameters, bolometric correction, and extinction reported by those authors. Combined with the GAIA DR3 parallax, these values imply a luminosity $\log L/L_\odot=4.08\pm 0.10$, similar to that of $\eta$~Leo. Again, this value differs significantly from those of \citet{martin18} and \citet{2021A&A...650A.128G} due to our use of the updated parallax value. The location of the star on the HR diagram is nearly coincident with $\eta$~Leo. The implied mass and radius yield a $\log g$ that is in good agreement with the measured value. 

Eight nights of observations spanning 4 years have been obtained for 13 Mon. The first, reported by \cite{martin18}, resulted in a ND. The more recent observations are of significantly higher S/N, leading to 7 profiles with visually apparent Zeeman signatures and flat null profiles (see Fig.~\ref{fig:lpv_etaleo13mon})\footnote{Two observations of this star - CFHT odometer numbers 2373414 (obtained on 27 January 2019) and 2378709 (obtained on 18 March 2019) - were corrupted, and were removed from the analysis.}. Integrating over a range of [$-9,33$]~kms, three of these signatures correspond to formal DDs, while one is a MD. The profiles with visible signatures are simple and antisymmetric, and exhibit a clear polarity change between November 2017 and January 2018. The longitudinal field (Fig.~\ref{Figbz}) varied from $2.61\pm 0.55$~G to $-1.90\pm 0.8$~G, exhibiting a change consistent with the polarity flip of the signatures. 

The spectrum of 13~Mon is essentially stable over the period of observation. No emission is present in H$\alpha$, nor is any present in the cores of the Ca~{\sc ii} H and K lines. As would be expected from their inferred atmospheric parameters the spectra of $\eta$~Leo and 13~Mon are very similar. The spectral lines of 13~Mon are somewhat narrower, in agreement with the results of \citet{2012A&A...543A..80F}.

\begin{figure}
\centering
\includegraphics[width=8cm]{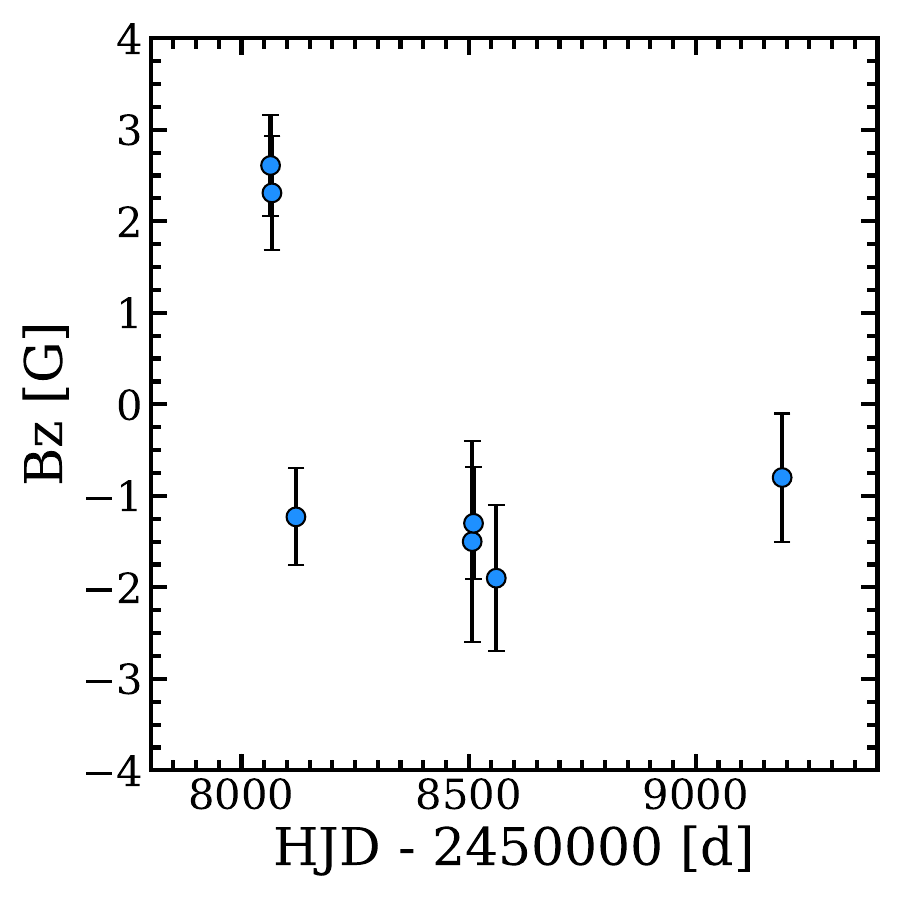}
\centering
\caption{Longitudinal field measurements of 13~Mon versus HJD. }
\label{fig:13monbz}
\end{figure}

As with $\eta$~Leo, given the simple morphology and stability of 13 Mon's Stokes $V$ signatures, and the star's high effective temperature, a fossil origin of the star's magnetic field could be plausible. The measured maximum longitudinal field of $|B_\ell|=2.61$~G would imply a flux-conservation-extrapolated ZAMS dipole of at least 1.2~kG.

Continued monitoring at similar precision with ideally monthly cadence for the next $\sim 5$ years is required to confirm the variability, establish its origins, and potentially determine the star's rotation period. 



\section{Discussion}\label{discuss}

This paper describes the persistent detection of Stokes $V$ Zeeman signatures in LSD profiles of four Iab/Ib supergiants spanning spectral types from F5 to A0.

\subsection{Magnetic field characteristics}

The instantaneous Stokes $V$ Zeeman signatures (see e.g. Fig.~\ref{figLSD}) of our targets range significantly in structural complexity, suggesting an analogous range of complexity of the underlying magnetic field morphologies. $\alpha$~Per (F5Iab, the coolest star of our sample) exhibits the most complex Stokes $V$ signatures, often with 4 changes of polarity across its profile. The Stokes $V$ profiles become systematically less complex with increasing spectral type, with the earliest stars (13 Mon and $\eta$~Leo, with respective spectral types of A1Ib and A0Ib) consistently exhibiting the simplest forms.

All targets show temporal variability of their Stokes $V$ profiles (see Figs. \ref{fig:lpv_aper}, \ref{fig:lpv_alep}, \ref{fig:lpv_etaleo13mon}), including both variations in complexity and changes in polarity accompanied by reversal of sign of the longitudinal field. The least variable star is $\eta$~Leo, whose profiles were non-variable from 2017-2019, until in 2024 the profiles appeared to be undergoing a polarity change. $\alpha$~Per exhibited the most rapid changes, with clear variability in Stokes $V$ structure on timescales of a few weeks, leading to the detection of periodic variability on a timescale of 50.75~d or 90~d.

The unsigned longitudinal magnetic fields measured from the LSD Stokes profiles (see Table~\ref{tab:Ultimate Table}) range up to about 2.5~G. By any measure these are globally weak magnetic fields, comparable in strength to those detected in late-type dwarfs \citep{2014MNRAS.444.3517M} and red giants \citep[e.g.][]{2021A&A...646A.130A}.  Comparing to the fields detected in the A supergiants $\iota$ Car (A7Ib), $\upsilon$~Car (A7Ib; \citealt{neiner17}), and 19~Aur (A5Ib-II; \citealt{martin18}), $\iota$~Car and $\upsilon$~Car are intermediate in temperature ($T_{\rm eff}\simeq 7500$~K) between the cooler and hotter stars of our sample. The Stokes $V$ profiles of $\iota$~Car appear simple, with a single polarity change across the profiles, and vary significantly in amplitude (but not in overall polarity) over $\sim 120$~days in 2014/15. The unsigned longitudinal field varied from 0.3 to 0.9~G (with typical uncertainties of about 0.2~G). Those of $\upsilon$~Car are considerably noisier, but also suggest a simple structure with no significant variability over the short (4 nights) span of the observations. The longitudinal field was measured between 1 and 2~G, with uncertainties of about 1~G. 19 Aur \citep[with $T_{eff} = 8500$~K;][]{martin18} also exhibits simple Zeeman signatures in its Stokes $V$ LSD profiles.  Some variability is seen, however, primarily only the positive pole is visible. The timescale of the variability appears to be on the order of several years \citep{2021mobs.confE..47O}. Our sample is shown in comparison to $\iota$ Car, $\upsilon$~Car, and 19~Aur in the HR diagram of Fig.~\ref{Fig:HR diagram}.


\subsection{Magnetic field origins}

Throughout the course of this study, we have attempted to draw conclusions about the physical processes or scenarios responsible for the magnetic fields that we have detected. Our argumentation has been based principally on the morphology and variability of the observed Zeeman signatures: complex Zeeman signatures, characterized by multiple polarity changes of the circular polarization across the LSD profile, are ascribed to magnetic fields of more complex morphology, more probably of dynamo origin. On the other hand, simple Zeeman signatures, characterized by a single Stokes $V$ polarity change across the LSD profile, coupled with secular polarity change of the overall Stokes $V$ profile (implying a simple, likely dipolar, field topology), are interpreted as potentially of fossil origin. \citet{neiner17} and \citet{martin18} used a similar rationale, and informed by stellar structure predictions from model evolutionary calculations proposed that the fields of $\iota$ Car, $\upsilon$~Car, and 19~Aur were remnants of fossil fields from the main sequence. 

In our sample, the hotter stars $\eta$~Leo and $13$~Mon exhibit Zeeman signatures most similar to those of these 3 stars: simple profiles exhibiting rather slow variability, qualitatively consistent with simple, organized surface magnetic fields observationally modulation by stellar rotation. In contrast, the cooler stars of our sample - $\alpha$~Per and $\alpha$~Lep - show behavior that is obviously different: complex Zeeman signatures, varying (at least in the case of $\alpha$~Per) rather rapidly. We would venture that while we may not be able to ascribe with full confidence the magnetic fields of the hotter stars to fossil fields, the cooler stars very likely have predominantly dynamo-generated fields. However, the situation may not be so simple: as pointed out by e.g. \citet{neiner17}, the low gravities of tepid supergiants may result in superficial convections zones, potentially permitting both a dynamo field and a fossil field simultaneously. Local dynamos may develop in the convective zones that appear as the star evolves and likely interact with the fossil field. The interaction can enhance the local dynamo and modify the configuration of the fossil field \cite[e.g.][]{2009ApJ...705.1000F,2008A&A...491..499A}. 

This behaviour is qualitatively consistent with the Stokes $V$ profiles of other cool supergiants. For example, Betelgeuse \citep[M1-M2Ia/Iab;][]{2018A&A...615A.116M} produces complex and variable Zeeman signatures attributed to a highly structured magnetic field generated by a dynamo powered by giant convection cells. Polaris \citep[][and Barron in prep.; F7-F8Ib]{barron22} exhibits Zeeman signatures that are also similar - in both morphology and variability - to those of $\alpha$~Per. In fact, \citet{2024IAUS..361..233B} reports that in a magnetic survey of classical Cepheid variables (all cool supergiants with spectral types ranging from F5Iab to G5Iab), 8 of 15 show clear evidence of Zeeman signatures. As was similarly concluded by \citet{grunhut10}, this rate is considerably higher than would be expected for fossil fields given our understanding of main sequence stars, which exhibit roughly a 10\% fraction of magnetic stars in this mass range \citep{grunhut17}.

This rationale, while founded on high-resolution spectropolarimetric observations of hundreds of stars spanning the HR diagram, is admittedly prone to ambiguity. For example, some stars reasonably interpreted to host dynamos (e.g. fully-convective M-dwarfs) exhibit spectropolarimetric characteristics that could well be interpreted to imply fossil fields under our argumentation above. Moreover, the origin of the magnetic fields of Vega and other ``ultra-weakly magnetic" A-type dwarfs (which arguably bear some spectropolarimetric similarity to the supergiants discussed here and by \citealt{barron22}) remains poorly understood. Hence spectropolarimetric behaviour should be combined with and supported by other properties to ensure accurate conclusions.

One potentially distinguishing characteristic is the presence of a chromosphere or corona. In Section~\ref{results} we discuss the absence of any clear emission in the Ca~{\sc ii} H and K lines or H$\alpha$ (potentially diagnostic of a chromosphere) in our sample.  \citet{2017ApJ...837...14A} reported that $\alpha$~Per was detected in X-rays as a hard coronal source ($T\sim 10$~MK) in a 1993 pointing by ROSAT. This result was confirmed by \citet{2018ApJ...854...95A}, who also reported significant X-ray count rates for $\alpha$~Lep and $\iota$~Car. While F-type supergiants generally exhibit peculiar UV and X-ray properties as compared to cooler G-type supergiants \citep{2018ApJ...854...95A}, there seems to be general consensus that these stars have chromospheres and coronae, and therefore presumably dynamo-powered magnetic fields.

As discussed by \cite{neiner17}, stellar structure models may provide some further insight into field origins, since an absence of convection near the stellar surface could suggest an inefficient dynamo capability. However, the limitations of currently evolutionary models (consideration of surface and interior magnetic fields, as well as historical binary interaction) imply that detailed (likely MHD) simulations will likely be required for a more fulsome understanding. 

Ultimately, further monitoring to better establish field variability characteristics and timescales is necessary if firmer conclusions are to be drawn about the nature of the magnetic fields of supergiants. Furthermore, a significant expansion of the sample is required to better understand the systematics and statistics of supergiant magnetism.

\section*{Acknowledgments}
This research has made use of the SIMBAD database operated at CDS, Strasbourg (France), NASA's Astrophysics Data System (ADS) and the Canadian Astronomy Data Centre (CADC). This work has made use of the VALD database, operated at Uppsala University, the Institute of Astronomy RAS in Moscow, and the University of Vienna. GAW acknowledges Discovery Grant support from the Natural Sciences and Engineering Research Council (NSERC) of Canada. MEO gratefully acknowledges support for this  work from the National Science Foundation under Grant No. AST-2107871. JAB acknowledges support from an NSERC Postgraduate Doctoral Scholarship (PGS D).

\section*{Data availability}
All spectropolarimetric data discussed in this paper are available via the Polarbase database ({\tt polarbase.irap.omp.eu}).

\bibliography{Paper}

\end{document}